\documentclass[twoside, epsfig]{article}

\input oejv.sty

\usepackage{longtable}
\usepackage{amssymb}

\newcommand{\dotem}{\rlap{.}^m}

\def\OEJVtitleTwo[#1]#2 {\uppercase
  {\centerline {\ftitle #1} \vskip 0.3cm
   \centerline {\ftitle #2} \vskip 0.5cm plus 0.4cm minus 0.2cm}}

\begin{document}

\OEJVhead{xxx 2025}
\OEJVtitleTwo[THE HADS STAR CSS\_J102714.3+205943:]{A COMPONENT OF A BINARY SYSTEM WITH AN ELLIPTIC ORBIT?}
\OEJVauth{Pyatnytskyy, Maksym Yu.$^1$ and Andronov, Ivan L.$^2$} 
\OEJVinst{ Private Observatory ”Osokorky”, PO Box 27, 02132 Kyiv, Ukraine, {\tt \href{mailto:pmak@osokorky-observatory.com}{pmak@osokorky-observatory.com}}}
\OEJVinst{ Department of Mathematics, Physics and Astronomy, Odesa National Maritime University, Mechnikova 34,
65029 Odesa, Ukraine} 

\OEJVabstract{We analyzed period changes of the high-amplitude Delta Scuti variable star CSS\_J102714.3+205943 for about 20 years, utilizing data from the automated sky surveys along with our own observations. With the help of the $O-C$ diagram, we found that the period decreased noticeably between JD2454800 and JD2457300. A possible cause of the change could be intrinsic processes in the star. However, the observed behavior of the $O-C$ diagram can also be explained by the light-time effect if the star is a component of a binary system. Times of maxima for the star, derived from the surveys and our observations, are listed.}

\begintext

\section{Introduction}\label{secintro} 

As defined in \citep{baglin1973}, Delta Scuti stars are short-period (from 0.02 to 0.2 days) pulsating variables of spectral types between A2V and F0V. They occupy a place in the Hertzsprung-Russell diagram where the extension of the Cepheid instability strip crosses the main sequence \citep{breger2000a, handler2009}. A more recent definition of the Delta Scuti variability type (DSCT) from the International Variable Star Index (VSX) \citep{watson2006} is broader and includes "pulsating variables of spectral types A0-F5 III-V displaying light amplitudes from 0.003 to 0.9 magnitudes in $V$" \footnote{\OEJVlink{https://www.aavso.org/vsx/index.php?view=about.vartypes}}, with periods from 0.01 to 0.2 days. 
The subset of Delta Scuti stars with a relatively high range of pulsations above $0\dotem3$ in $V$ \citep{breger2000b, handler2009} ($0\dotem15$ according to VSX) are called high-amplitude DSCT stars (HADS). 

Delta Scuti stars are important for asteroseismology \citep{breger2000a}. Also, because they have their own period-luminosity relationship \citep{mcnamara1997, mcnamara2011, ziaali2019}, these stars can be used as 'standard candles' for measuring distances \citep{mcnamara2000}.

\begin{figure}[htbp]
\centering
\includegraphics[width=14cm]{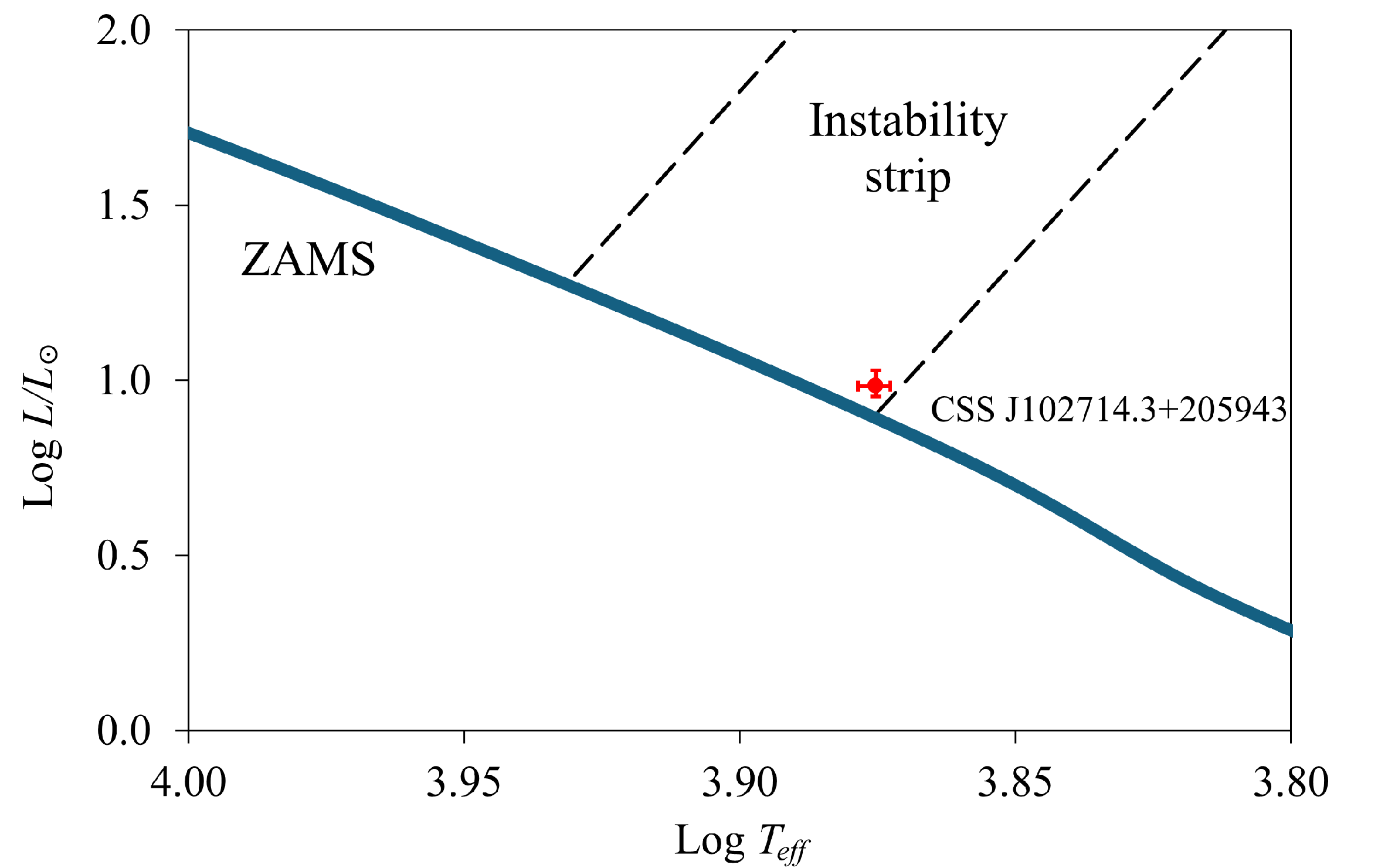} 
\caption{The Hertzsprung–Russell diagram with the Zero Age Main Sequence (ZAMS) and the instability strip. The red dot with error bars indicates the location of CSS\_J102714.3+205943 according to the Gaia DR3 catalogue.}
\label{FigHRD}
\end{figure}

In the current research, we investigated the period change of the poorly studied HADS star CSS\_J102714.3+205943 = GSC 01426-00590 = ZTF J102714.30+205943.4 = TIC 171599792. According to VSX, the period of the star is 0.0684195 days, the variability range is $0 \dotem 442$ in the $g$ band. Using the estimated star's luminosity ($9.6 L_\odot$)and effective temperature ($7505 K$) from \citep{gaia2022}, we can see that the star is indeed located in the area of the HR diagram where the instability strip crosses the main sequence (Fig.~\ref{FigHRD}).

\section{Observation and data reduction}

We analyzed our observations and data from automatic surveys. For primary data manipulation, such as converting between different survey formats, visually inspecting, and removing outliers, we utilized the VStar software \citep{benn2012, benn2024}.

\subsection{Our observations}

The observation of CSS\_J102714.3+205943 was carried out in February 2022 from Kyiv, Ukraine, for three nights. We used a 150mm Newtonian telescope with a focal length of 750mm, equipped with the cooled monochrome CMOS camera ZWO ASI183MM Pro and the $V$ photometric filter.

The light frames were treated with a standard procedure using dark, flat, and dark-flat calibration frames. The calibration frames were taken every observation night immediately after the observing session. To take the flat frames, we used a white paper screen placed about 2 meters from the telescope illuminated by an LED lamp.

\begin{figure}[htbp]
\centering
\includegraphics[width=14cm]{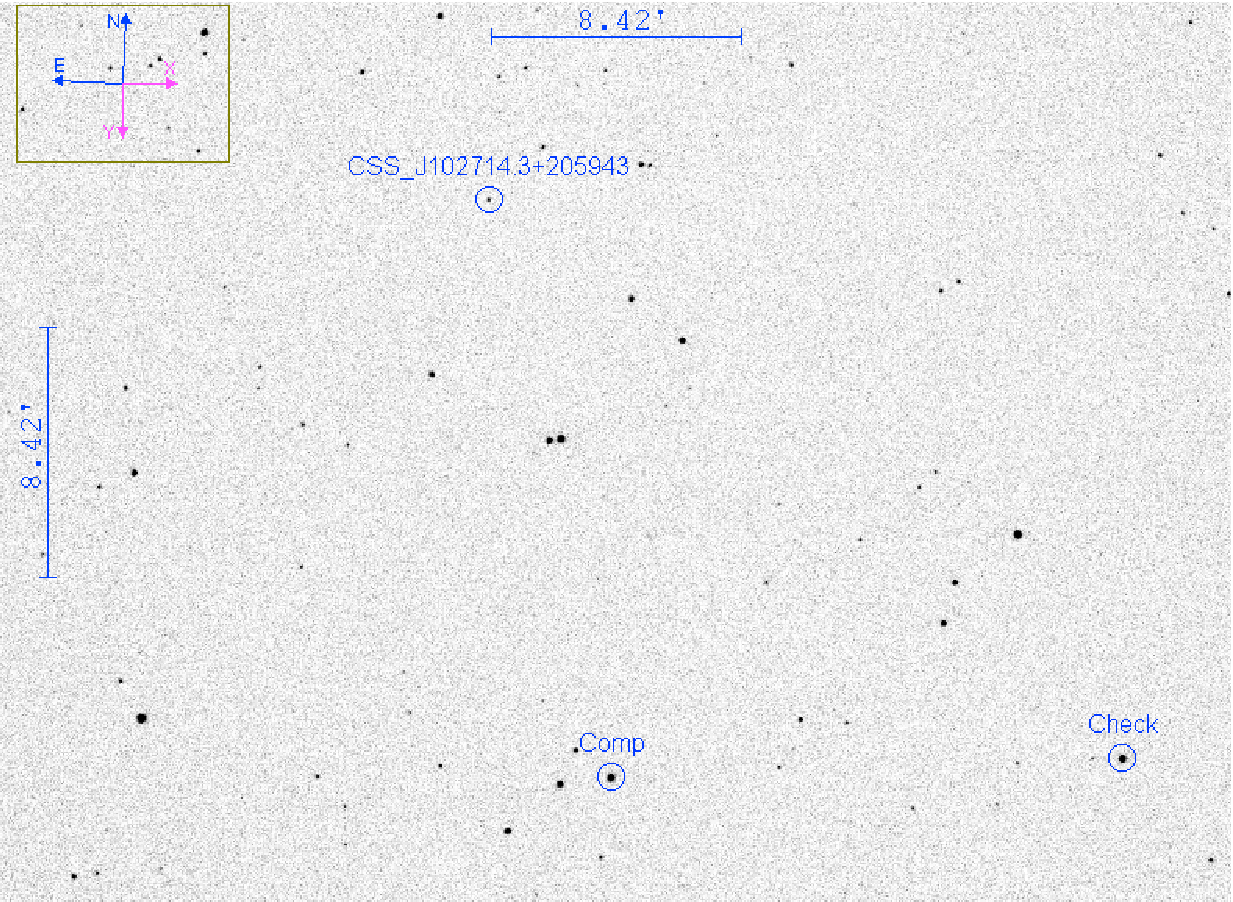} 
\caption{A sample image showing the FOV with the object and the comparison stars.}
\label{FigFOV-n}
\end{figure}

A sample image with the object and the comparison stars is shown in Fig.~\ref{FigFOV-n}. These stars were selected according to the AAVSO recommended sequence\footnote{\OEJVlink{https://apps.aavso.org/vsp/photometry/?chartid=X38581CXQ}}. Magnitudes of the stars were taken from the APASS DR10 survey \citep{henden2018}. Detailed information about the comparison stars is given in Tab.~\ref{tabCompStars}.

The differential aperture photometry was conducted using the AstroImageJ software \citep{collins2017}.

\begin{figure}[htbp]
\centering
\includegraphics[width=14cm]{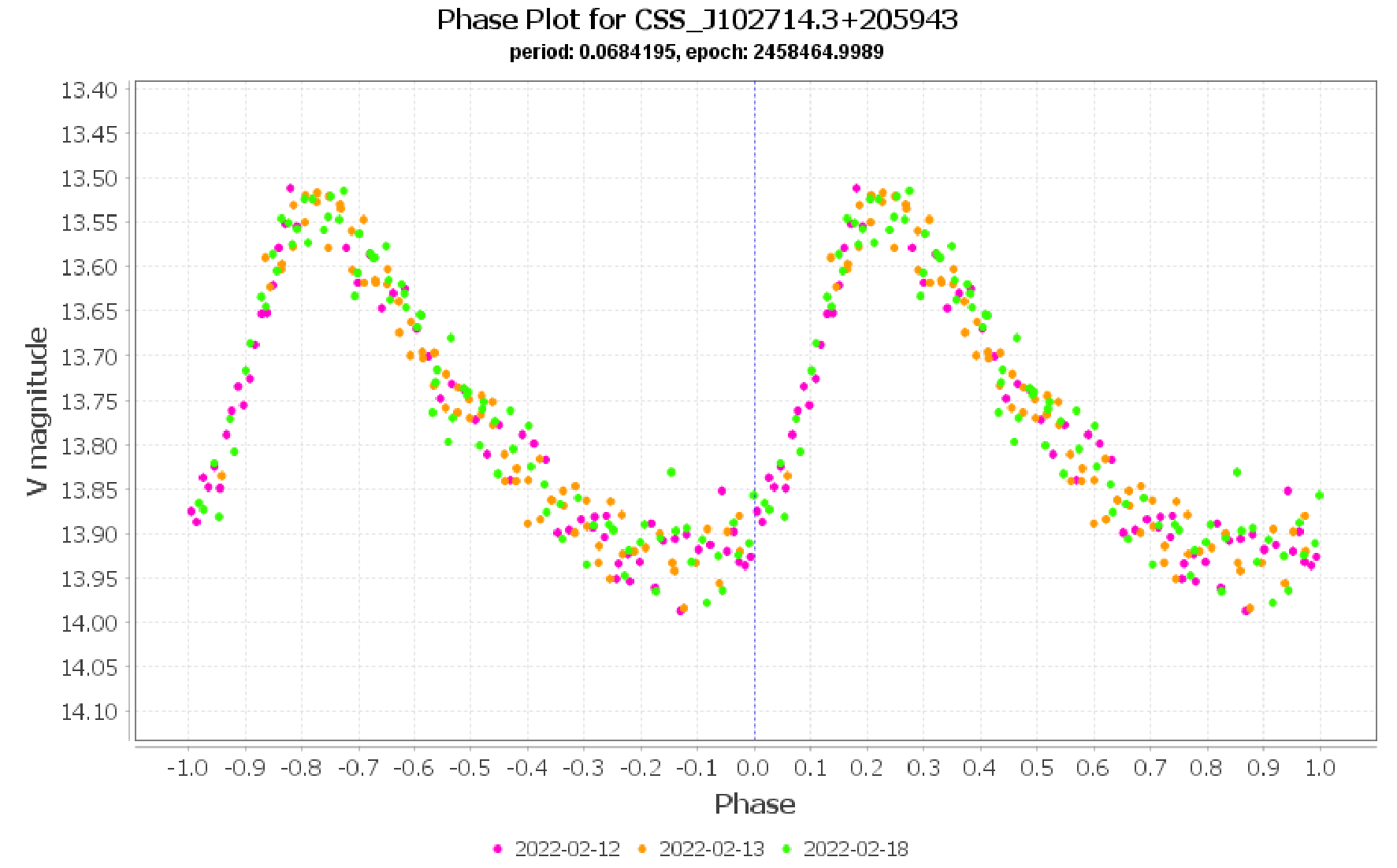} 
\caption{Phase plot of our observations over three nights (see legend). The period and initial epoch are taken from VSX for November 2024.}
\label{figPMAK_PhasePlot}
\end{figure}

A phase plot of the resulting light curve is shown in Fig.~\ref{figPMAK_PhasePlot}.

\begin{table} 
\caption{Comparison and check stars.}\vspace{3mm}  
\centering
\begin{tabular}{llccr}
\hline
	 ID & Type & RA (2000) [h:m:s] & DEC (2000) [$^{\circ}$ $'$ $''$] & $V$ [mag]\\ \hline \hline
  TYC 1426-1013-1 & Comp  & 10:26:52.9 & +20:40:29.3 & 10.784(69) \\
  TYC 1426-979-1  & Check & 10:25:39.6 & +20:41:52.8 & 10.999(69) \\
\hline
\end{tabular}\label{tabCompStars}
\end{table}

\subsection{Data from the TESS survey}

Another source of the continuous light curve was the TESS survey \citep{ricker2014}.

Checking the MAST Portal\footnote{\OEJVlink{https://mast.stsci.edu/portal/Mashup/Clients/Mast/Portal.html}} revealed that there were no ready-to-use light curves available. However, there were TESS Full Frame Images (FFI) data for TESS sectors 45, 46, 48, and 72. The exposure lengths for sectors 45, 46, and 48 were too long (475 seconds) for the star with such a short period. The exposure for sector 72 was 158 seconds (with 200 seconds between frames), which is appropriate.

We used the 'Lightkurve' Python package \citep{lightkurve2018} to generate the TESS light curve. 

\begin{figure}[htbp]
\centering
\includegraphics[width=14cm]{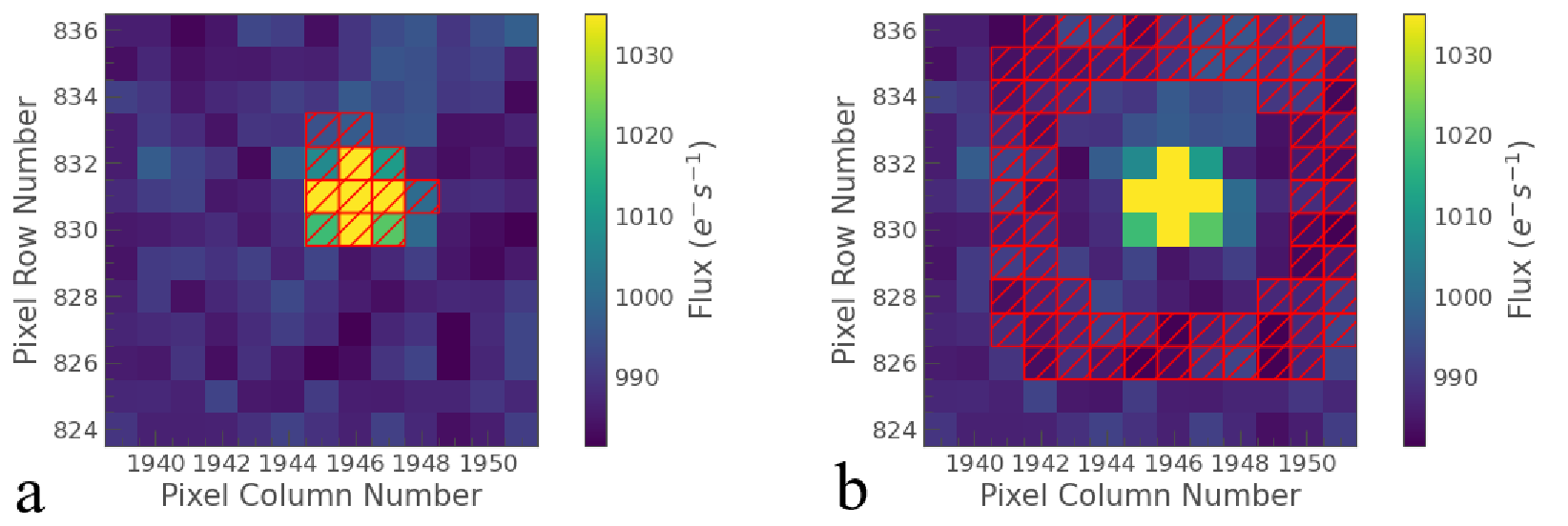} 
\caption{Apertures used for calculating the star's flux from TESS target pixel images: (a) the measurement aperture; (b) the aperture used for estimating the background signal level.}
\label{FigTESS-aperture}
\end{figure}

\begin{figure}[htbp]
\centering
\includegraphics[width=14cm]{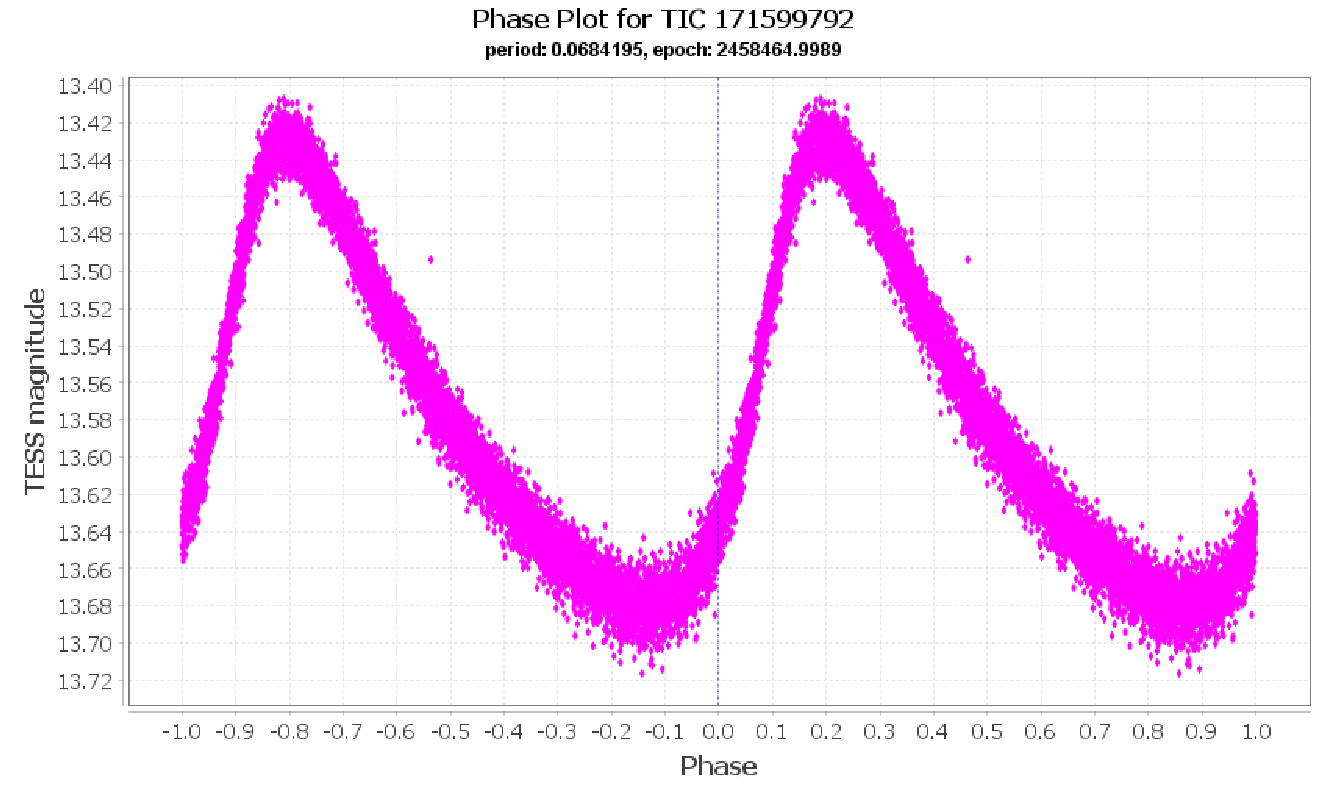} 
\caption{TESS phase plot. The period and initial epoch are taken from VSX for November 2024.}
\label{figTESS_PhasePlot}
\end{figure}

For sector 72, we extracted a set of target pixel images (TPI) of the star's vicinity of size $13 \times 13$ pixels, corresponding to $4.55 \times 4.55$ arcminutes. Then, we created a measurement aperture with the 'create\_threshold\_mask' function, taking pixels with a median flux that is greater than 7 times the standard deviation above the overall median. A sample target image with the aperture mask is shown in Fig.~\ref{FigTESS-aperture}a.
To estimate the background level, we used the background aperture shown in Fig.~\ref{FigTESS-aperture}b. 
For each TPI, we calculated the star's flux inside the measurement aperture, subtracted the estimated background level, and thus obtained the light curve. Then, we visually inspected the light curve and rejected parts with apparent parasitic flux. Using the TESS magnitude of the star from the TESS Input Catalog - v8.0 \citep{stassun2019} ($13\dotem579$) as the reference mean magnitude, we converted the fluxes to magnitudes, obtaining the final light curve. A phase plot of the TESS light curve is shown in Fig.~\ref{figTESS_PhasePlot}.

\subsection{Data from other surveys}

For our analysis, we also used data from other surveys, namely, CRTS \citep{drake2009}, SuperWASP \citep{butters2010}, ASAS-SN \citep{kochanek2017}, and ZTF \citep{masci2019}.

\section{Data analysis}\label{secdata}

When we tried to combine all the data, we found that it was impossible to choose a single period for all observations. Therefore, we assumed that the period changes over time. A commonly used method for investigating period changes is the $O-C$ diagram. To build the diagram, we need precise times of maxima (TOMs).

To derive TOMs from the data, we first built a model light curve based on our observations in the $V$ filter, approximating it with trigonometric polynomials. We used the approach described in \citep{andronov1994, andronov2020} to determine the statistically optimal approximation, which corresponds to the minimal r.m.s. error of the approximations. The following formula defines the optimal approximation:

\begin{equation}\label{eq1}
\sigma_m^2[x_C]=\frac{m}{n(n-m)}\sum_{k=1}^n (x_k - x_{Ck})^2
\end{equation}

where $n$ is the number of observations, $m$ is the number of parameters, $x_k$ is a $k$-th observation, $x_{Ck}$ is a $k$-th calculated (model) point. The number of parameters $m$ is equal to $2s+2$, where $s$ is the degree of the polynomial (the number of harmonics, including the primary frequency). Another free parameter is the constant level, and an additional one comes from adjusting the primary frequency.

With this approach, utilizing the MCV software \citep{andronov2004, andronov2020}, we determined that, besides the main frequency $f_0 = 1 / P_0$ (where $P_0 = 0.0684223(33)$ days), we need the second harmonic ($2 \cdot f_0$) and the third harmonic ($3 \cdot f_0$) to adequately approximate the observations, i.e. the $\sigma_m^2[x_C]$ value minimizes for $s=3$.

\begin{figure}[htbp]
\centering
\includegraphics[width=14cm]{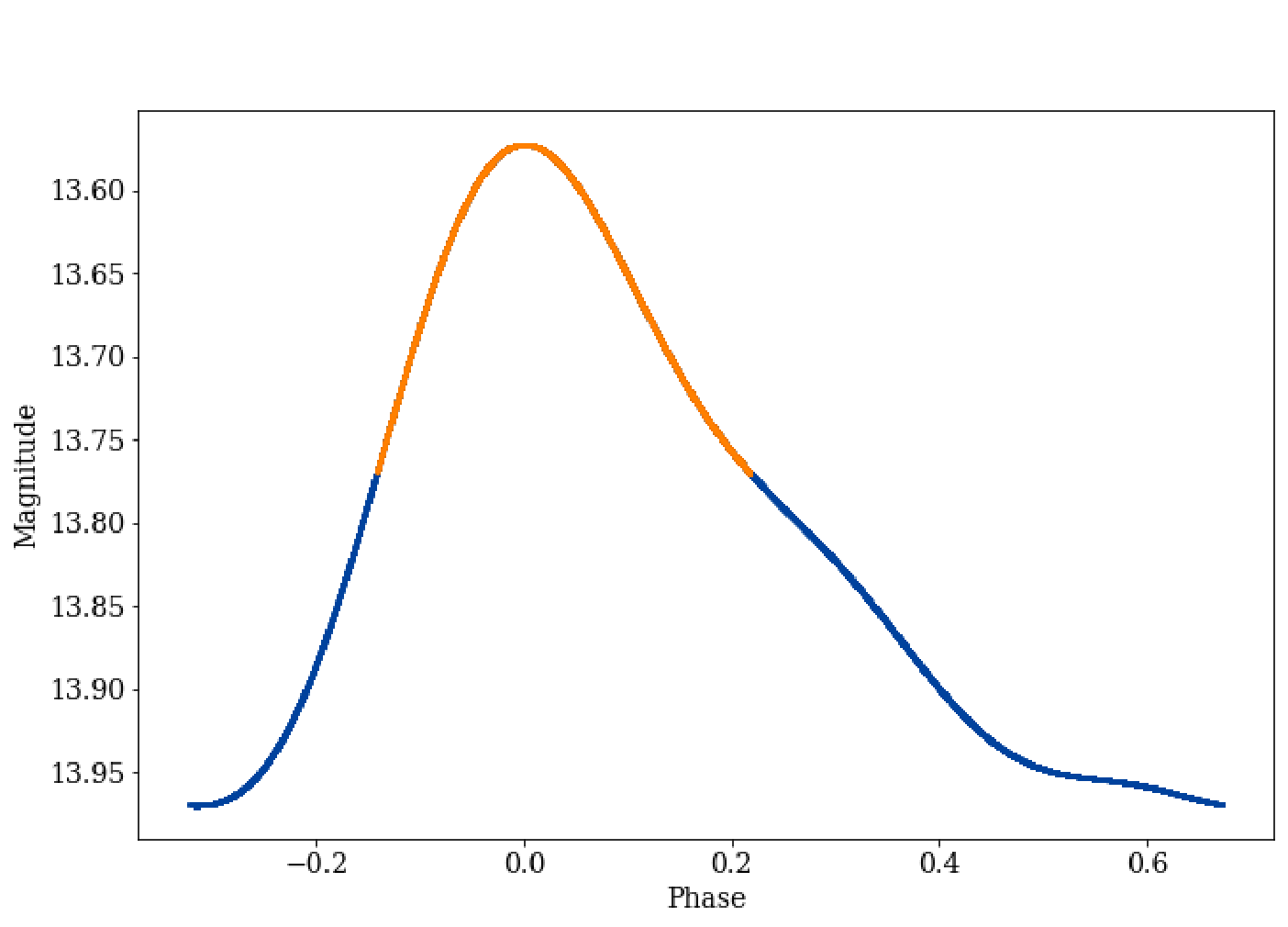} 
\caption{The light curve model for our observations was built using a trigonometric polynomial of the third order. The upper part of the model (light colored) was used for the approximation of TOMs.}
\label{figLC_Model}
\end{figure}

The resulting light curve model is shown in Fig.~\ref{figLC_Model}. We used the upper part of the model (from the top to half of the maximum) to fit the observed light curve near the maxima, thereby finding the TOMs. We successfully utilized this approach previously for another HADS star \citep{pyatnytskyy2021, pyatnytskyy2024}.

We used this model for our observations and data from the surveys except for TESS. Given that the survey data are sparse, we folded the survey data over a long period of time before determining the times of maxima (see Tab.~\ref{tabOC}, column "Half of the folding interval").

\begin{figure}[htbp]
\centering
\includegraphics[width=14cm]{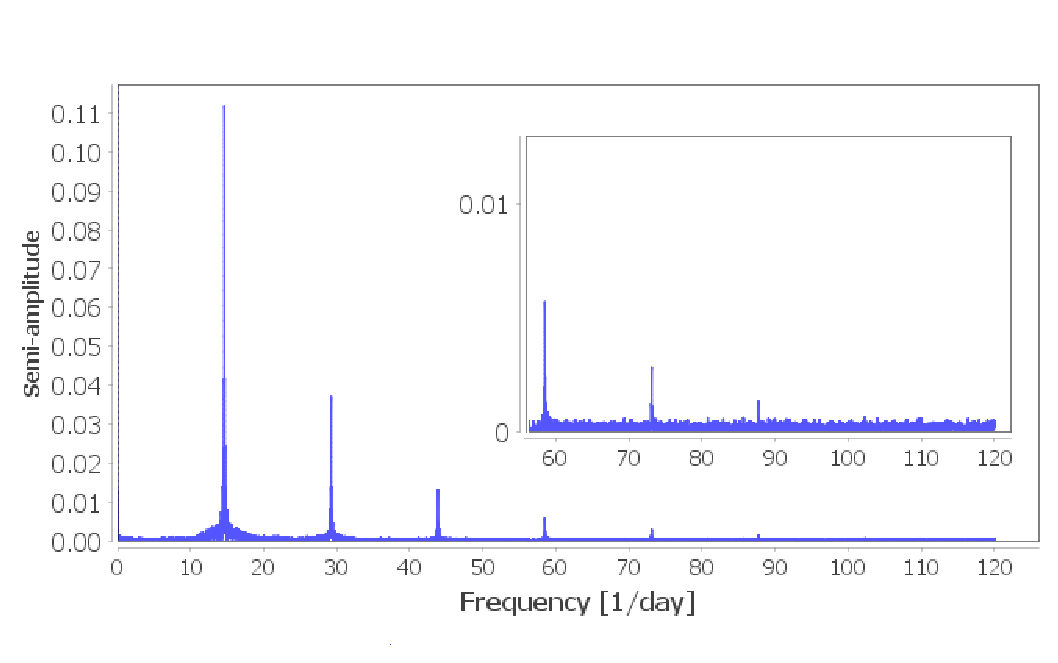} 
\caption{Periodogram of the TESS data.}
\label{FigTESS-periodogram}
\end{figure}

We built a separate model for TESS data due to the smaller variability range in the TESS band and the somewhat different curve shape. In this case, we used six harmonics (including the primary frequency) for the approximation. We chose this number of harmonics based on the periodogram (created with VStar) of the TESS data, see Fig.~\ref{FigTESS-periodogram}.

We found no additional frequencies in the TESS light curve except for the primary one and its harmonics. This suggests that the star is a radial pulsator with a single pulsation mode. This is not unexpected, because although Delta Scuti stars, in general, may exhibit many superimposed pulsation modes \citep{breger2000b}, there is evidence that most HADS stars pulsate in a single mode \citep{Xue2023}.

\pagebreak
\section{Results}\label{secresults}

Tab.~\ref{tabOC} lists TOMs derived from all the data.

\begin{figure}[htbp]
\centering
\includegraphics[width=14cm]{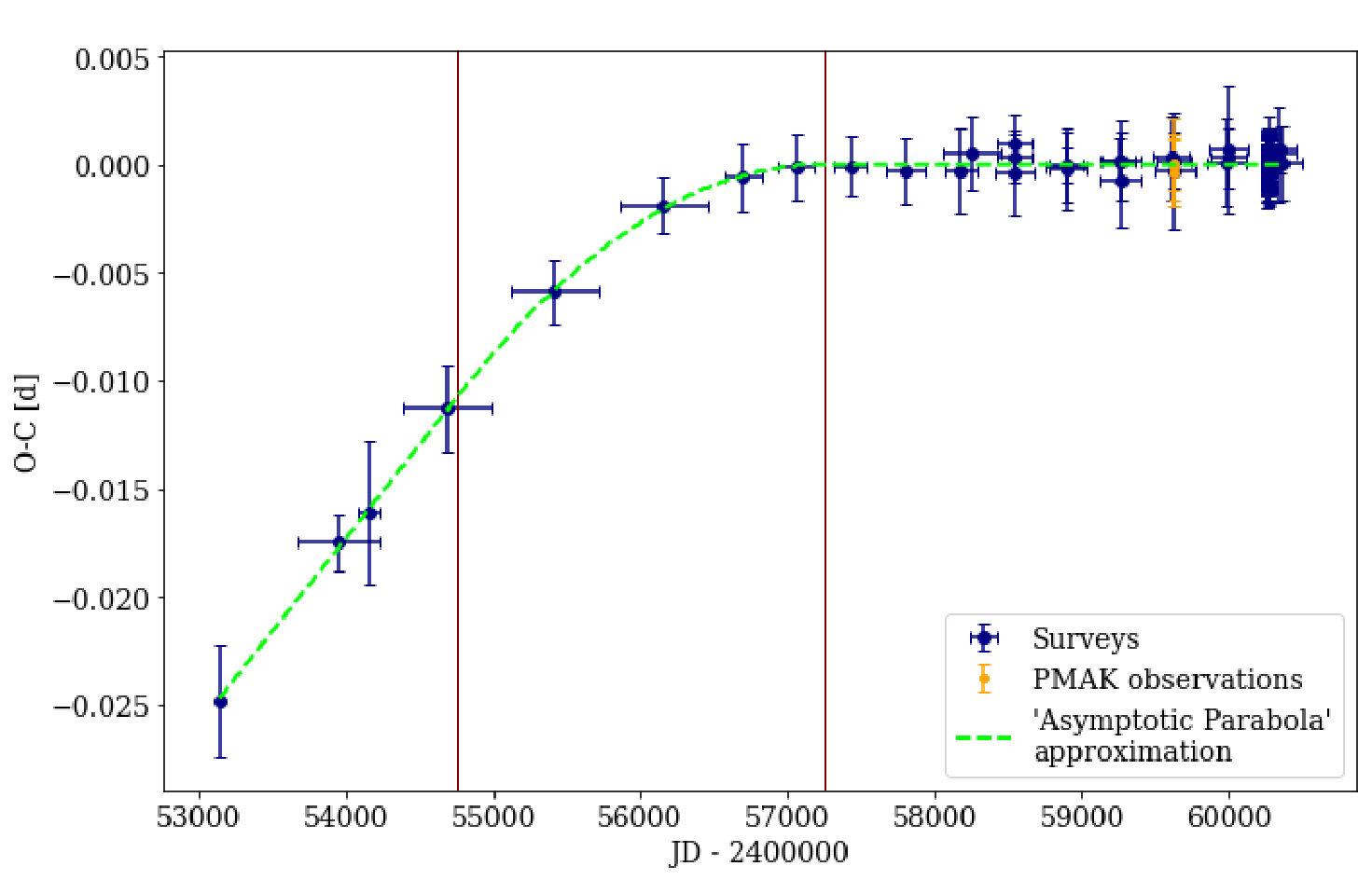} 
\caption{$O-C$ diagram corresponding to Eq.~\ref{eq2}. Our observations are labeled as 'PMAK'. Horizontal error bars show folding intervals for the data from the surveys. Vertical lines denote the AP approximation's parabolic part's start and end.}
\label{FigO-C}
\end{figure}

Having determined the TOMs, we built the $O-C$ diagram, as shown in Fig.~\ref{FigO-C}. The diagram corresponds to the following ephemeris:

\begin{equation}\label{eq2}
T_{max}[BJD_{TDB}]=2459623.42653(3)+0.068419093(3)E
\end{equation}

where $T_{max}$ is the time of a maximum for a cycle number $E$.

The period in Eq.~\ref{eq2} corresponds to the horizontal flat part of the $O-C$ diagram that starts from $JD \approx 2457300$. This value comes from the approximation using the Asymptotic Parabola (AP) method \citep{andrych2015} realized in the MAVKA software \citep{andrych2020} \footnote{\OEJVlink{https://uavso.org.ua/mavka/}}. 

From the AP approximation, the ascending straight branch of the $O-C$ diagram ends at $JD \approx 2454800$, and the horizontal flat part starts at the previously mentioned value of $JD \approx 2457300$, with a smooth transition between these points. With this interpretation of the $O-C$ diagram, we can state that the star's period decreases from a constant value of 0.068419686(11) to 0.068419093(3) $d^{-1}$.

Similar period changes have already been observed in other stars of the HADS type \citep{laney2003, axelsen2020}. Their nature, however, is still unclear.

\begin{figure}[htbp]
\centering
\includegraphics[width=14cm]{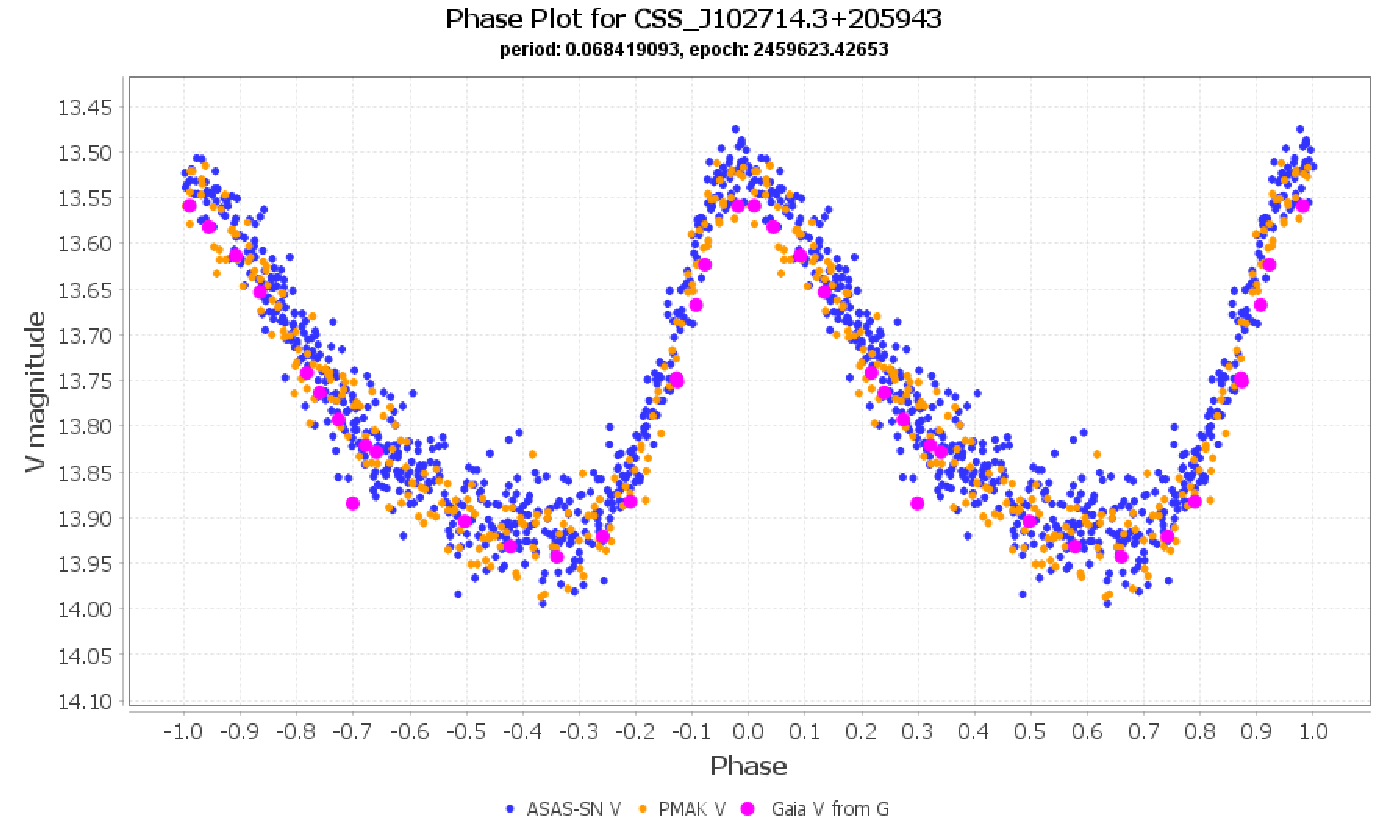} 
\caption{Phase plot with our observations (PMAK), ASAS-SN data in the $V$ filter, and Gaia DR3 data transformed to the $V$ band. The ASAS-SN and Gaia DR3 data are for $JD > 2457300$. The period and initial epoch are from Eq.~\ref{eq2}.}
\label{FigV_PhasePlot_2457300}
\end{figure}

Fig.~\ref{FigV_PhasePlot_2457300} demonstrates the phased light curve for $JD > 2457300$, which includes datasets in the $V$ filter, i.e., our observations and a part of ASAS-SN data. We also added available photometric data from the Gaia DR3 release \citep{gaia2023} to this plot. There are only 29 Gaia DR3 observations in total (each in the $G$, $BP$, and $RP$ filters), and only 20 of them since JD2457300 to date, so we did not use them while creating the $O-C$ diagram. Nevertheless, the Gaia DR3 data are in good agreement with other observations. An average Gaia DR3 $V$ magnitude is somewhat dimmer than the one from our observations and ASAS-SN data by $\approx0\dotem03$. This can be due to uncertainties in the transformation of the $G$, $BP$, and $RP$ values to $V$. See the corresponding relationship in the 'Photometric relationships with other photometric systems' section here \footnote{\OEJVlink{https://gea.esac.esa.int/archive/documentation/GDR2/}}. From the phase plot, we derived the previously unknown variability range in the $V$ filter and the light curve rise duration time from minimum to maximum magnitude, which characterizes the light curve's asymmetry (see Tab.~\ref{tabLCparam}).

\begin{table} 
\caption{Light curve parameters in the $V$ filter for $JD > 2457300$.}\vspace{3mm}  
\centering
\begin{tabular}{lc}
\hline
	 Parameter & Value\\ \hline \hline
  Magnitude at minimum & 13.92 \\
  Magnitude at maximum & 13.52 \\
  Period [d] & 0.068419093 \\
  Initial epoch [$BJD_{TDB}$] & 2459623.4265 \\
  Rise duration [\%] & 32 \\
\hline
\end{tabular}\label{tabLCparam}
\end{table}

However, another interpretation of this bent $O-C$ diagram could be considered. If the star is a component of a binary system, the light-time effect could cause the apparent period changes \citep{sterken2005}. With this assumption, the observed deviations in the times of maxima are caused by the star's shift along the line of sight. If a hypothetical second body orbits around the common center of mass, the observed star also moves in an elliptical orbit around the center of mass.

Using the MCV software, we approximated the $O-C$ curve with a second-order trigonometric polynomial with a linear trend. This gave us a starting point for further approximation. Then, we used a combination of manual fitting of the model's parameters and approximation with the OCFit package \citep{gaidos2019, gaidos2023}. The interval of the observations is too short and does not cover a single complete orbit. Therefore, for now, we intend to show only the possibility of this interpretation without expecting a precise result.

\begin{figure}[htbp]
\centering
\includegraphics[width=14cm]{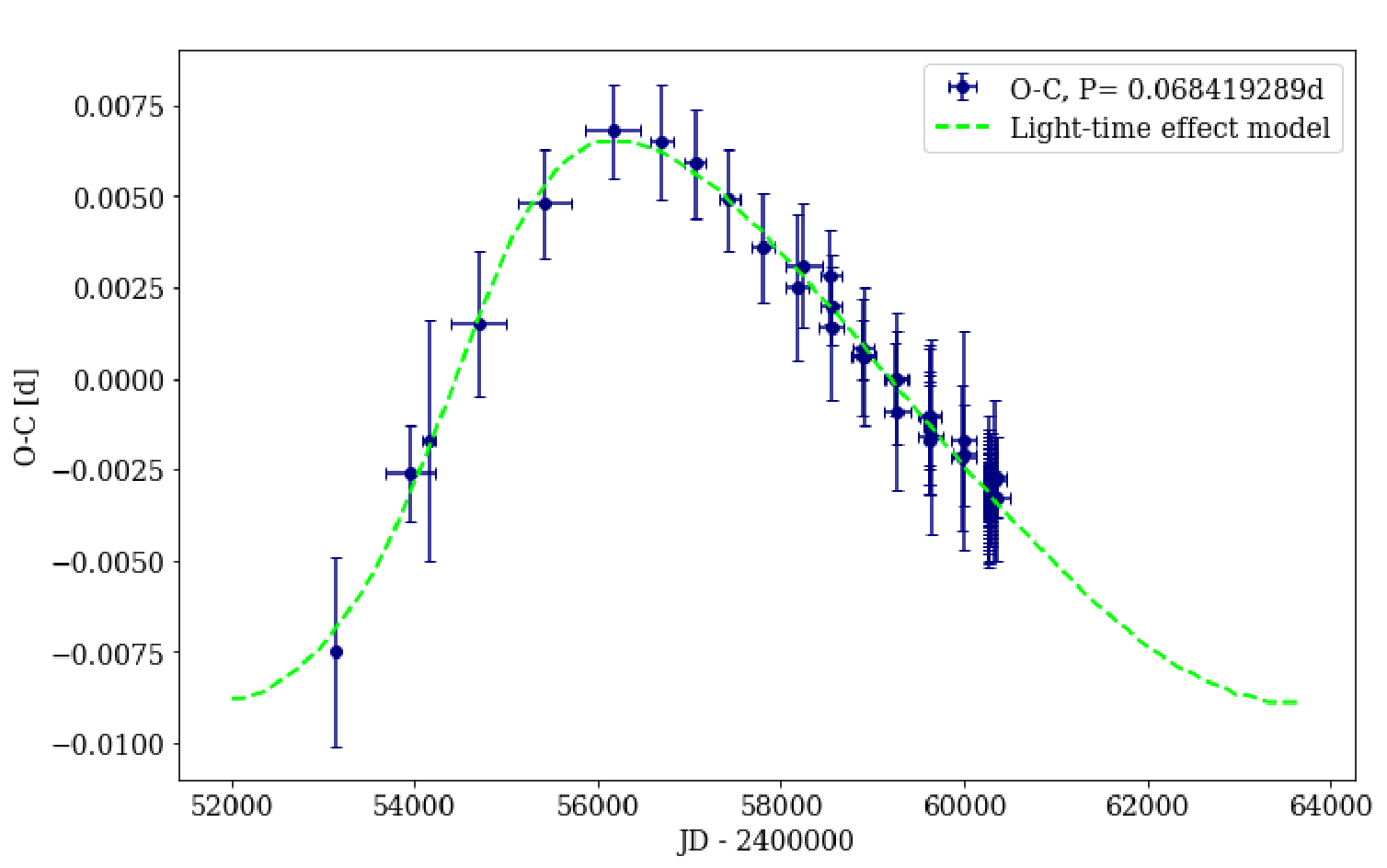} 
\caption{Modified $O-C$ diagram with the light-time effect model.}
\label{FigO-C_LightTime}
\end{figure}

Fig.~\ref{FigO-C_LightTime} shows the result of the modeling. We ended up with the following parameters of the model: rotation period is $32.3 yrs$, the value of $a\sin i$ is $1.45a.u.$ (where $a$ is the orbit's semi-major axis, $i$ is the orbit's inclination), orbit's eccentricity is $0.41$, and argument of pericenter is $20^{\circ}$. Further observations over several dozen years are required to prove our interpretation.

If this interpretation is valid, we can estimate the lower limit of the second body's mass. Given that the mass of the observed HADS star CSS\_J102714.3+205943 is $1.66 M_\odot$ \citep{gaia2022}, the estimated mass of the second body is $\gtrsim 0.2 M_\odot$. So, the second body is likely a red dwarf.

\section{Conclusions}\label{secconc}

We found a period change in the HADS star CSS\_J102714.3+205943, which occurred between JD2454800 and JD2457300. The observed change in the period could be caused by intrinsic processes. However, a more straightforward interpretation is possible if we assume that the star is a component of a binary system. If this assumption is valid, the apparent change in the period is caused by a light-time effect. The second component of the binary is likely a red dwarf with a lower mass limit of about $0.2 M_\odot$. Further long-term observations are necessary to prove this assumption.

\setcounter{secnumdepth}{0}
\OEJVacknowledgements{
This paper includes results derived from data collected by the TESS mission. Funding for the TESS mission is provided by the NASA's Science Mission Directorate.\\
This work makes use of observations obtained with the Samuel Oschin Telescope 48-inch and the 60-inch Telescope at the Palomar
Observatory as part of the Zwicky Transient Facility project. ZTF is supported by the National Science Foundation under Grant
No. AST-2034437 and a collaboration including Caltech, IPAC, the Weizmann Institute for Science, the Oskar Klein Center at
Stockholm University, the University of Maryland, Deutsches Elektronen-Synchrotron and Humboldt University, the TANGO
Consortium of Taiwan, the University of Wisconsin at Milwaukee, Trinity College Dublin, Lawrence Livermore National
Laboratories, and IN2P3, France. Operations are conducted by COO, IPAC, and UW.\\
We thank the Las Cumbres Observatory and its staff for their continuing support of the ASAS-SN project.\\
This paper makes use of data from the DR1 of the WASP data \citep{butters2010} as provided by the WASP consortium, and computational resources supplied by the project "e-Infrastruktura CZ" (e-INFRA CZ LM2018140) supported by the Ministry of Education, Youth and Sports of the Czech Republic.\\
We make use of the data from the CSS survey, which is funded by the National Aeronautics and Space
Administration under Grant No. NNG05GF22G issued through the Science
Mission Directorate Near-Earth Objects Observations Program.  The CRTS
survey is supported by the U.S.~National Science Foundation under
grants AST-0909182.
} \\

\begin{longtable}[c]{ccccccc}
\caption{Times of maxima and $O-C$ values calculated for the period and initial epoch specified in Tab.~\ref{tabLCparam}.\label{tabOC}}
\\ \hline
$BJD_{TDB}$ & Error & Epoch          & $O-C$ & Half of the      & Source & Filter \\
 $-2400000$ & [d]   & [cycle number] & [d] & folding interval &        &        \\
            &       &                &     & [d]              &        &        \\
\hline \hline
53138.4348 & 0.0026 & -94783 & -0.0248 & 8 & SuperWASP & Clear \\
54154.4671 & 0.0033 & -79933 & -0.0161 & 71 & SuperWASP & Clear \\
53952.3556 & 0.0013 & -82887 & -0.0175 & 279 & CRTS & Clear \\
54694.8459 & 0.0020 & -72035 & -0.0113 & 304 & CRTS & Clear \\
55418.8621 & 0.0015 & -61453 & -0.0059 & 300 & CRTS & Clear \\
56159.3660 & 0.0013 & -50630 & -0.0019 & 294 & CRTS & Clear \\
58180.7412 & 0.0020 & -21086 & -0.0003 & 117 & ASAS-SN & g \\
58546.0991 & 0.0020 & -15746 & -0.0004 & 128 & ASAS-SN & g \\
58906.6679 & 0.0019 & -10476 & -0.0002 & 129 & ASAS-SN & g \\
59268.2624 & 0.0022 & -5191 & -0.0007 & 137 & ASAS-SN & g \\
59634.2365 & 0.0027 & 158 & -0.0003 & 135 & ASAS-SN & g \\
59998.7743 & 0.0030 & 5486 & 0.0007 & 136 & ASAS-SN & g \\
60365.2948 & 0.0017 & 10843 & 0.0001 & 134 & ASAS-SN & g \\
56701.9307 & 0.0016 & -42700 & -0.0006 & 126 & ASAS-SN & V \\
57060.4472 & 0.0015 & -37460 & -0.0001 & 122 & ASAS-SN & V \\
57429.9788 & 0.0014 & -32059 & -0.0001 & 116 & ASAS-SN & V \\
57807.3098 & 0.0015 & -26544 & -0.0003 & 130 & ASAS-SN & V \\
58251.0768 & 0.0017 & -20058 & 0.0005 & 197 & ASAS-SN & V \\
58546.3734 & 0.0011 & -15742 & 0.0003 & 116 & ZTF & g \\
58895.3789 & 0.0008 & -10641 & 0.0000 & 118 & ZTF & g \\
59254.3742 & 0.0010 & -5394 & 0.0002 & 113 & ZTF & g \\
59625.4109 & 0.0013 & 29 & 0.0002 & 115 & ZTF & g \\
59998.2950 & 0.0014 & 5479 & 0.0003 & 118 & ZTF & g \\
60357.8377 & 0.0011 & 10734 & 0.0007 & 107 & ZTF & g \\
58544.8689 & 0.0013 & -15764 & 0.0010 & 118 & ZTF & r \\
58899.3471 & 0.0016 & -10583 & -0.0001 & 137 & ZTF & r \\
59266.8265 & 0.0018 & -5212 & 0.0002 & 137 & ZTF & r \\
59613.8482 & 0.0019 & -140 & 0.0003 & 125 & ZTF & r \\
59986.3899 & 0.0020 & 5305 & 0.0001 & 130 & ZTF & r \\
60341.8275 & 0.0022 & 10500 & 0.0005 & 121 & ZTF & r \\
59623.4265 & 0.0015 & 0 & -0.0001 & n.a. & PMAK & V \\
59624.3840 & 0.0015 & 14 & -0.0004 & n.a. & PMAK & V \\
59629.3107 & 0.0020 & 86 & 0.0001 & n.a. & PMAK & V \\
59629.4475 & 0.0012 & 88 & 0.0000 & n.a. & PMAK & V \\
60262.3241 & 0.0012 & 9338 & 0.0001 & n.a. & TESS & TESS \\
60262.3925 & 0.0010 & 9339 & 0.0000 & n.a. & TESS & TESS \\
60262.4605 & 0.0007 & 9340 & -0.0003 & n.a. & TESS & TESS \\
60262.5293 & 0.0014 & 9341 & 0.0000 & n.a. & TESS & TESS \\
60262.5976 & 0.0007 & 9342 & 0.0000 & n.a. & TESS & TESS \\
60262.6663 & 0.0012 & 9343 & 0.0002 & n.a. & TESS & TESS \\
60262.7344 & 0.0013 & 9344 & -0.0002 & n.a. & TESS & TESS \\
60262.8025 & 0.0012 & 9345 & -0.0004 & n.a. & TESS & TESS \\
60262.8716 & 0.0008 & 9346 & 0.0002 & n.a. & TESS & TESS \\
60262.9396 & 0.0010 & 9347 & -0.0002 & n.a. & TESS & TESS \\
60263.0086 & 0.0012 & 9348 & 0.0004 & n.a. & TESS & TESS \\
60263.0773 & 0.0009 & 9349 & 0.0007 & n.a. & TESS & TESS \\
60263.1451 & 0.0008 & 9350 & 0.0001 & n.a. & TESS & TESS \\
60263.2134 & 0.0010 & 9351 & -0.0001 & n.a. & TESS & TESS \\
60263.2820 & 0.0011 & 9352 & 0.0001 & n.a. & TESS & TESS \\
60263.3507 & 0.0007 & 9353 & 0.0004 & n.a. & TESS & TESS \\
60263.4184 & 0.0016 & 9354 & -0.0004 & n.a. & TESS & TESS \\
60263.4868 & 0.0005 & 9355 & -0.0003 & n.a. & TESS & TESS \\
60263.5554 & 0.0007 & 9356 & -0.0002 & n.a. & TESS & TESS \\
60263.6240 & 0.0013 & 9357 & 0.0001 & n.a. & TESS & TESS \\
60263.6918 & 0.0011 & 9358 & -0.0006 & n.a. & TESS & TESS \\
60263.7609 & 0.0008 & 9359 & 0.0000 & n.a. & TESS & TESS \\
60263.8289 & 0.0017 & 9360 & -0.0003 & n.a. & TESS & TESS \\
60263.8976 & 0.0011 & 9361 & 0.0000 & n.a. & TESS & TESS \\
60263.9661 & 0.0008 & 9362 & 0.0000 & n.a. & TESS & TESS \\
60264.0348 & 0.0014 & 9363 & 0.0003 & n.a. & TESS & TESS \\
60264.1028 & 0.0005 & 9364 & -0.0001 & n.a. & TESS & TESS \\
60264.1713 & 0.0006 & 9365 & -0.0001 & n.a. & TESS & TESS \\
60264.2403 & 0.0010 & 9366 & 0.0005 & n.a. & TESS & TESS \\
60264.3076 & 0.0009 & 9367 & -0.0005 & n.a. & TESS & TESS \\
60264.3764 & 0.0013 & 9368 & -0.0001 & n.a. & TESS & TESS \\
60264.4450 & 0.0013 & 9369 & 0.0000 & n.a. & TESS & TESS \\
60264.5133 & 0.0009 & 9370 & -0.0001 & n.a. & TESS & TESS \\
60264.5816 & 0.0008 & 9371 & -0.0003 & n.a. & TESS & TESS \\
60264.6503 & 0.0015 & 9372 & 0.0001 & n.a. & TESS & TESS \\
60264.7188 & 0.0012 & 9373 & 0.0001 & n.a. & TESS & TESS \\
60264.7872 & 0.0010 & 9374 & 0.0001 & n.a. & TESS & TESS \\
60264.8552 & 0.0010 & 9375 & -0.0003 & n.a. & TESS & TESS \\
60264.9243 & 0.0011 & 9376 & 0.0004 & n.a. & TESS & TESS \\
60264.9922 & 0.0007 & 9377 & -0.0002 & n.a. & TESS & TESS \\
60265.0610 & 0.0008 & 9378 & 0.0002 & n.a. & TESS & TESS \\
60265.1292 & 0.0007 & 9379 & 0.0000 & n.a. & TESS & TESS \\
60265.1977 & 0.0012 & 9380 & 0.0000 & n.a. & TESS & TESS \\
60265.2663 & 0.0012 & 9381 & 0.0002 & n.a. & TESS & TESS \\
60265.3341 & 0.0013 & 9382 & -0.0004 & n.a. & TESS & TESS \\
60265.4029 & 0.0007 & 9383 & 0.0000 & n.a. & TESS & TESS \\
60265.4711 & 0.0008 & 9384 & -0.0002 & n.a. & TESS & TESS \\
60265.5392 & 0.0010 & 9385 & -0.0005 & n.a. & TESS & TESS \\
60265.6078 & 0.0008 & 9386 & -0.0004 & n.a. & TESS & TESS \\
60265.6761 & 0.0009 & 9387 & -0.0005 & n.a. & TESS & TESS \\
60265.7450 & 0.0011 & 9388 & 0.0000 & n.a. & TESS & TESS \\
60265.8135 & 0.0008 & 9389 & 0.0001 & n.a. & TESS & TESS \\
60265.8817 & 0.0007 & 9390 & -0.0002 & n.a. & TESS & TESS \\
60265.9501 & 0.0009 & 9391 & -0.0001 & n.a. & TESS & TESS \\
60266.0185 & 0.0011 & 9392 & -0.0002 & n.a. & TESS & TESS \\
60266.0868 & 0.0012 & 9393 & -0.0003 & n.a. & TESS & TESS \\
60266.1552 & 0.0008 & 9394 & -0.0003 & n.a. & TESS & TESS \\
60266.2238 & 0.0010 & 9395 & -0.0001 & n.a. & TESS & TESS \\
60266.2926 & 0.0010 & 9396 & 0.0003 & n.a. & TESS & TESS \\
60266.6347 & 0.0007 & 9401 & 0.0003 & n.a. & TESS & TESS \\
60266.7027 & 0.0007 & 9402 & -0.0001 & n.a. & TESS & TESS \\
60266.7713 & 0.0012 & 9403 & 0.0000 & n.a. & TESS & TESS \\
60266.8399 & 0.0012 & 9404 & 0.0002 & n.a. & TESS & TESS \\
60266.9077 & 0.0009 & 9405 & -0.0003 & n.a. & TESS & TESS \\
60266.9764 & 0.0011 & 9406 & -0.0002 & n.a. & TESS & TESS \\
60267.0449 & 0.0008 & 9407 & 0.0000 & n.a. & TESS & TESS \\
60267.1130 & 0.0008 & 9408 & -0.0003 & n.a. & TESS & TESS \\
60267.1821 & 0.0007 & 9409 & 0.0003 & n.a. & TESS & TESS \\
60267.2501 & 0.0008 & 9410 & -0.0001 & n.a. & TESS & TESS \\
60267.3186 & 0.0010 & 9411 & 0.0000 & n.a. & TESS & TESS \\
60267.3872 & 0.0014 & 9412 & 0.0002 & n.a. & TESS & TESS \\
60267.4550 & 0.0010 & 9413 & -0.0005 & n.a. & TESS & TESS \\
60267.5238 & 0.0009 & 9414 & 0.0000 & n.a. & TESS & TESS \\
60267.5920 & 0.0014 & 9415 & -0.0003 & n.a. & TESS & TESS \\
60267.6607 & 0.0006 & 9416 & 0.0000 & n.a. & TESS & TESS \\
60267.7289 & 0.0006 & 9417 & -0.0002 & n.a. & TESS & TESS \\
60267.7975 & 0.0005 & 9418 & -0.0001 & n.a. & TESS & TESS \\
60267.8664 & 0.0010 & 9419 & 0.0005 & n.a. & TESS & TESS \\
60267.9343 & 0.0010 & 9420 & -0.0001 & n.a. & TESS & TESS \\
60268.0026 & 0.0012 & 9421 & -0.0002 & n.a. & TESS & TESS \\
60268.0711 & 0.0008 & 9422 & -0.0002 & n.a. & TESS & TESS \\
60268.1397 & 0.0009 & 9423 & 0.0000 & n.a. & TESS & TESS \\
60268.2078 & 0.0004 & 9424 & -0.0003 & n.a. & TESS & TESS \\
60268.2762 & 0.0008 & 9425 & -0.0003 & n.a. & TESS & TESS \\
60268.3452 & 0.0005 & 9426 & 0.0003 & n.a. & TESS & TESS \\
60268.4134 & 0.0011 & 9427 & 0.0001 & n.a. & TESS & TESS \\
60268.4821 & 0.0010 & 9428 & 0.0003 & n.a. & TESS & TESS \\
60268.5501 & 0.0007 & 9429 & -0.0001 & n.a. & TESS & TESS \\
60268.6183 & 0.0013 & 9430 & -0.0003 & n.a. & TESS & TESS \\
60268.6872 & 0.0008 & 9431 & 0.0002 & n.a. & TESS & TESS \\
60268.7560 & 0.0010 & 9432 & 0.0006 & n.a. & TESS & TESS \\
60268.8238 & 0.0007 & 9433 & 0.0000 & n.a. & TESS & TESS \\
60268.8924 & 0.0010 & 9434 & 0.0002 & n.a. & TESS & TESS \\
60268.9608 & 0.0009 & 9435 & 0.0001 & n.a. & TESS & TESS \\
60269.0293 & 0.0015 & 9436 & 0.0002 & n.a. & TESS & TESS \\
60269.0977 & 0.0007 & 9437 & 0.0002 & n.a. & TESS & TESS \\
60269.1657 & 0.0008 & 9438 & -0.0002 & n.a. & TESS & TESS \\
60269.2346 & 0.0008 & 9439 & 0.0002 & n.a. & TESS & TESS \\
60269.3030 & 0.0007 & 9440 & 0.0002 & n.a. & TESS & TESS \\
60269.3713 & 0.0012 & 9441 & 0.0001 & n.a. & TESS & TESS \\
60269.4395 & 0.0010 & 9442 & -0.0001 & n.a. & TESS & TESS \\
60269.5081 & 0.0008 & 9443 & 0.0001 & n.a. & TESS & TESS \\
60269.5765 & 0.0013 & 9444 & 0.0000 & n.a. & TESS & TESS \\
60269.6450 & 0.0011 & 9445 & 0.0001 & n.a. & TESS & TESS \\
60269.7136 & 0.0018 & 9446 & 0.0004 & n.a. & TESS & TESS \\
60269.7816 & 0.0004 & 9447 & -0.0001 & n.a. & TESS & TESS \\
60269.8500 & 0.0005 & 9448 & -0.0002 & n.a. & TESS & TESS \\
60269.9189 & 0.0008 & 9449 & 0.0004 & n.a. & TESS & TESS \\
60269.9872 & 0.0009 & 9450 & 0.0003 & n.a. & TESS & TESS \\
60270.0560 & 0.0010 & 9451 & 0.0006 & n.a. & TESS & TESS \\
60270.1240 & 0.0013 & 9452 & 0.0002 & n.a. & TESS & TESS \\
60270.1922 & 0.0006 & 9453 & 0.0000 & n.a. & TESS & TESS \\
60270.2609 & 0.0006 & 9454 & 0.0003 & n.a. & TESS & TESS \\
60270.3288 & 0.0011 & 9455 & -0.0003 & n.a. & TESS & TESS \\
60270.3977 & 0.0010 & 9456 & 0.0002 & n.a. & TESS & TESS \\
60270.4660 & 0.0006 & 9457 & 0.0001 & n.a. & TESS & TESS \\
60270.5344 & 0.0007 & 9458 & 0.0001 & n.a. & TESS & TESS \\
60270.6025 & 0.0010 & 9459 & -0.0002 & n.a. & TESS & TESS \\
60270.6710 & 0.0007 & 9460 & -0.0001 & n.a. & TESS & TESS \\
60270.7393 & 0.0010 & 9461 & -0.0003 & n.a. & TESS & TESS \\
60270.8080 & 0.0008 & 9462 & 0.0000 & n.a. & TESS & TESS \\
60270.8767 & 0.0012 & 9463 & 0.0003 & n.a. & TESS & TESS \\
60270.9446 & 0.0010 & 9464 & -0.0002 & n.a. & TESS & TESS \\
60271.0134 & 0.0007 & 9465 & 0.0001 & n.a. & TESS & TESS \\
60271.0819 & 0.0007 & 9466 & 0.0002 & n.a. & TESS & TESS \\
60271.1503 & 0.0010 & 9467 & 0.0002 & n.a. & TESS & TESS \\
60271.2186 & 0.0005 & 9468 & 0.0001 & n.a. & TESS & TESS \\
60271.2868 & 0.0006 & 9469 & -0.0001 & n.a. & TESS & TESS \\
60271.3553 & 0.0009 & 9470 & -0.0001 & n.a. & TESS & TESS \\
60271.4235 & 0.0006 & 9471 & -0.0002 & n.a. & TESS & TESS \\
60271.4918 & 0.0016 & 9472 & -0.0003 & n.a. & TESS & TESS \\
60271.5606 & 0.0008 & 9473 & 0.0000 & n.a. & TESS & TESS \\
60271.6286 & 0.0006 & 9474 & -0.0004 & n.a. & TESS & TESS \\
60271.6972 & 0.0006 & 9475 & -0.0002 & n.a. & TESS & TESS \\
60271.7661 & 0.0004 & 9476 & 0.0002 & n.a. & TESS & TESS \\
60271.8342 & 0.0008 & 9477 & -0.0001 & n.a. & TESS & TESS \\
60271.9026 & 0.0011 & 9478 & -0.0001 & n.a. & TESS & TESS \\
60271.9712 & 0.0011 & 9479 & 0.0001 & n.a. & TESS & TESS \\
60272.0394 & 0.0009 & 9480 & -0.0001 & n.a. & TESS & TESS \\
60272.1076 & 0.0012 & 9481 & -0.0004 & n.a. & TESS & TESS \\
60272.1760 & 0.0006 & 9482 & -0.0004 & n.a. & TESS & TESS \\
60272.2450 & 0.0009 & 9483 & 0.0002 & n.a. & TESS & TESS \\
60272.3129 & 0.0010 & 9484 & -0.0003 & n.a. & TESS & TESS \\
60272.3814 & 0.0007 & 9485 & -0.0002 & n.a. & TESS & TESS \\
60272.4497 & 0.0007 & 9486 & -0.0003 & n.a. & TESS & TESS \\
60272.5185 & 0.0009 & 9487 & 0.0001 & n.a. & TESS & TESS \\
60272.5868 & 0.0012 & 9488 & 0.0000 & n.a. & TESS & TESS \\
60272.6551 & 0.0009 & 9489 & -0.0002 & n.a. & TESS & TESS \\
60272.7237 & 0.0010 & 9490 & -0.0001 & n.a. & TESS & TESS \\
60272.7922 & 0.0010 & 9491 & 0.0000 & n.a. & TESS & TESS \\
60272.8607 & 0.0012 & 9492 & 0.0001 & n.a. & TESS & TESS \\
60272.9291 & 0.0015 & 9493 & 0.0001 & n.a. & TESS & TESS \\
60275.8709 & 0.0017 & 9536 & -0.0001 & n.a. & TESS & TESS \\
60275.9392 & 0.0008 & 9537 & -0.0003 & n.a. & TESS & TESS \\
60276.0080 & 0.0010 & 9538 & 0.0001 & n.a. & TESS & TESS \\
60276.0762 & 0.0007 & 9539 & 0.0000 & n.a. & TESS & TESS \\
60276.1448 & 0.0008 & 9540 & 0.0001 & n.a. & TESS & TESS \\
60276.2131 & 0.0016 & 9541 & 0.0000 & n.a. & TESS & TESS \\
60276.2816 & 0.0009 & 9542 & 0.0000 & n.a. & TESS & TESS \\
60276.3502 & 0.0011 & 9543 & 0.0003 & n.a. & TESS & TESS \\
60276.4184 & 0.0011 & 9544 & 0.0001 & n.a. & TESS & TESS \\
60276.4868 & 0.0015 & 9545 & 0.0000 & n.a. & TESS & TESS \\
60276.5554 & 0.0012 & 9546 & 0.0002 & n.a. & TESS & TESS \\
60276.6235 & 0.0008 & 9547 & -0.0001 & n.a. & TESS & TESS \\
60276.6922 & 0.0009 & 9548 & 0.0001 & n.a. & TESS & TESS \\
60276.7606 & 0.0005 & 9549 & 0.0002 & n.a. & TESS & TESS \\
60276.8289 & 0.0008 & 9550 & 0.0000 & n.a. & TESS & TESS \\
60276.8973 & 0.0009 & 9551 & 0.0000 & n.a. & TESS & TESS \\
60276.9657 & 0.0008 & 9552 & 0.0000 & n.a. & TESS & TESS \\
60277.0344 & 0.0009 & 9553 & 0.0003 & n.a. & TESS & TESS \\
60277.1026 & 0.0011 & 9554 & 0.0001 & n.a. & TESS & TESS \\
60277.1708 & 0.0009 & 9555 & -0.0001 & n.a. & TESS & TESS \\
60277.2395 & 0.0007 & 9556 & 0.0001 & n.a. & TESS & TESS \\
60277.3076 & 0.0010 & 9557 & -0.0002 & n.a. & TESS & TESS \\
60277.3759 & 0.0007 & 9558 & -0.0003 & n.a. & TESS & TESS \\
60277.4448 & 0.0012 & 9559 & 0.0002 & n.a. & TESS & TESS \\
60277.5128 & 0.0008 & 9560 & -0.0003 & n.a. & TESS & TESS \\
60277.5820 & 0.0009 & 9561 & 0.0005 & n.a. & TESS & TESS \\
60277.6498 & 0.0005 & 9562 & -0.0001 & n.a. & TESS & TESS \\
60277.7188 & 0.0007 & 9563 & 0.0005 & n.a. & TESS & TESS \\
60277.7864 & 0.0007 & 9564 & -0.0003 & n.a. & TESS & TESS \\
60277.8550 & 0.0007 & 9565 & -0.0002 & n.a. & TESS & TESS \\
60277.9238 & 0.0008 & 9566 & 0.0002 & n.a. & TESS & TESS \\
60277.9918 & 0.0011 & 9567 & -0.0001 & n.a. & TESS & TESS \\
60278.0605 & 0.0010 & 9568 & 0.0000 & n.a. & TESS & TESS \\
60278.1290 & 0.0014 & 9569 & 0.0001 & n.a. & TESS & TESS \\
60278.1970 & 0.0008 & 9570 & -0.0003 & n.a. & TESS & TESS \\
60278.2656 & 0.0008 & 9571 & -0.0001 & n.a. & TESS & TESS \\
60278.3339 & 0.0009 & 9572 & -0.0002 & n.a. & TESS & TESS \\
60278.4025 & 0.0005 & 9573 & 0.0000 & n.a. & TESS & TESS \\
60278.4708 & 0.0011 & 9574 & -0.0001 & n.a. & TESS & TESS \\
60278.5395 & 0.0009 & 9575 & 0.0001 & n.a. & TESS & TESS \\
60278.6079 & 0.0007 & 9576 & 0.0001 & n.a. & TESS & TESS \\
60278.6762 & 0.0013 & 9577 & 0.0000 & n.a. & TESS & TESS \\
60278.7447 & 0.0010 & 9578 & 0.0001 & n.a. & TESS & TESS \\
60278.8124 & 0.0009 & 9579 & -0.0006 & n.a. & TESS & TESS \\
60278.8813 & 0.0008 & 9580 & -0.0001 & n.a. & TESS & TESS \\
60278.9496 & 0.0006 & 9581 & -0.0002 & n.a. & TESS & TESS \\
60279.0182 & 0.0010 & 9582 & 0.0000 & n.a. & TESS & TESS \\
60279.0866 & 0.0008 & 9583 & -0.0001 & n.a. & TESS & TESS \\
60279.1555 & 0.0008 & 9584 & 0.0004 & n.a. & TESS & TESS \\
60279.2238 & 0.0008 & 9585 & 0.0003 & n.a. & TESS & TESS \\
60279.2918 & 0.0010 & 9586 & -0.0001 & n.a. & TESS & TESS \\
60279.5653 & 0.0009 & 9590 & -0.0003 & n.a. & TESS & TESS \\
60279.6342 & 0.0010 & 9591 & 0.0002 & n.a. & TESS & TESS \\
60279.7024 & 0.0011 & 9592 & -0.0001 & n.a. & TESS & TESS \\
60279.7708 & 0.0004 & 9593 & -0.0001 & n.a. & TESS & TESS \\
60279.8395 & 0.0010 & 9594 & 0.0002 & n.a. & TESS & TESS \\
60279.9075 & 0.0004 & 9595 & -0.0002 & n.a. & TESS & TESS \\
60279.9761 & 0.0009 & 9596 & 0.0000 & n.a. & TESS & TESS \\
60280.0448 & 0.0006 & 9597 & 0.0002 & n.a. & TESS & TESS \\
60280.1130 & 0.0008 & 9598 & 0.0000 & n.a. & TESS & TESS \\
60280.1815 & 0.0005 & 9599 & 0.0001 & n.a. & TESS & TESS \\
60280.2500 & 0.0005 & 9600 & 0.0002 & n.a. & TESS & TESS \\
60280.3186 & 0.0007 & 9601 & 0.0003 & n.a. & TESS & TESS \\
60280.3866 & 0.0008 & 9602 & -0.0001 & n.a. & TESS & TESS \\
60280.4552 & 0.0010 & 9603 & 0.0001 & n.a. & TESS & TESS \\
60280.5239 & 0.0005 & 9604 & 0.0004 & n.a. & TESS & TESS \\
60280.5920 & 0.0007 & 9605 & 0.0001 & n.a. & TESS & TESS \\
60280.6604 & 0.0005 & 9606 & 0.0001 & n.a. & TESS & TESS \\
60280.7284 & 0.0011 & 9607 & -0.0004 & n.a. & TESS & TESS \\
60280.7969 & 0.0013 & 9608 & -0.0003 & n.a. & TESS & TESS \\
60280.8656 & 0.0010 & 9609 & 0.0000 & n.a. & TESS & TESS \\
60280.9344 & 0.0012 & 9610 & 0.0004 & n.a. & TESS & TESS \\
60281.0025 & 0.0010 & 9611 & 0.0001 & n.a. & TESS & TESS \\
60281.0708 & 0.0011 & 9612 & -0.0001 & n.a. & TESS & TESS \\
60281.1390 & 0.0007 & 9613 & -0.0003 & n.a. & TESS & TESS \\
60281.2075 & 0.0012 & 9614 & -0.0002 & n.a. & TESS & TESS \\
60281.2761 & 0.0005 & 9615 & 0.0000 & n.a. & TESS & TESS \\
60281.3445 & 0.0009 & 9616 & -0.0001 & n.a. & TESS & TESS \\
60281.4127 & 0.0007 & 9617 & -0.0002 & n.a. & TESS & TESS \\
60281.4816 & 0.0008 & 9618 & 0.0002 & n.a. & TESS & TESS \\
60281.5496 & 0.0010 & 9619 & -0.0001 & n.a. & TESS & TESS \\
60281.6182 & 0.0010 & 9620 & 0.0000 & n.a. & TESS & TESS \\
60281.6867 & 0.0009 & 9621 & 0.0000 & n.a. & TESS & TESS \\
60281.7554 & 0.0009 & 9622 & 0.0003 & n.a. & TESS & TESS \\
60281.8236 & 0.0007 & 9623 & 0.0001 & n.a. & TESS & TESS \\
60281.8917 & 0.0012 & 9624 & -0.0001 & n.a. & TESS & TESS \\
60281.9600 & 0.0007 & 9625 & -0.0003 & n.a. & TESS & TESS \\
60282.0290 & 0.0011 & 9626 & 0.0003 & n.a. & TESS & TESS \\
60282.0970 & 0.0007 & 9627 & -0.0001 & n.a. & TESS & TESS \\
60282.1654 & 0.0007 & 9628 & -0.0002 & n.a. & TESS & TESS \\
60282.2338 & 0.0009 & 9629 & -0.0002 & n.a. & TESS & TESS \\
60282.3027 & 0.0012 & 9630 & 0.0003 & n.a. & TESS & TESS \\
60282.3709 & 0.0008 & 9631 & 0.0001 & n.a. & TESS & TESS \\
60282.4391 & 0.0005 & 9632 & -0.0001 & n.a. & TESS & TESS \\
60282.5076 & 0.0006 & 9633 & 0.0000 & n.a. & TESS & TESS \\
60282.5761 & 0.0008 & 9634 & 0.0000 & n.a. & TESS & TESS \\
60282.6444 & 0.0004 & 9635 & -0.0001 & n.a. & TESS & TESS \\
60282.7128 & 0.0010 & 9636 & -0.0001 & n.a. & TESS & TESS \\
60282.7821 & 0.0007 & 9637 & 0.0008 & n.a. & TESS & TESS \\
60282.8494 & 0.0008 & 9638 & -0.0004 & n.a. & TESS & TESS \\
60282.9183 & 0.0009 & 9639 & 0.0001 & n.a. & TESS & TESS \\
60282.9868 & 0.0007 & 9640 & 0.0002 & n.a. & TESS & TESS \\
60283.0548 & 0.0010 & 9641 & -0.0002 & n.a. & TESS & TESS \\
60283.1233 & 0.0009 & 9642 & -0.0001 & n.a. & TESS & TESS \\
60283.1920 & 0.0008 & 9643 & 0.0002 & n.a. & TESS & TESS \\
60283.2602 & 0.0011 & 9644 & 0.0000 & n.a. & TESS & TESS \\
60283.3288 & 0.0009 & 9645 & 0.0001 & n.a. & TESS & TESS \\
60283.3969 & 0.0011 & 9646 & -0.0002 & n.a. & TESS & TESS \\
60283.4653 & 0.0011 & 9647 & -0.0002 & n.a. & TESS & TESS \\
60283.5340 & 0.0011 & 9648 & 0.0000 & n.a. & TESS & TESS \\
60283.6027 & 0.0008 & 9649 & 0.0003 & n.a. & TESS & TESS \\
60283.6706 & 0.0009 & 9650 & -0.0002 & n.a. & TESS & TESS \\
60283.7392 & 0.0007 & 9651 & 0.0000 & n.a. & TESS & TESS \\
60283.8075 & 0.0011 & 9652 & -0.0001 & n.a. & TESS & TESS \\
60283.8758 & 0.0007 & 9653 & -0.0003 & n.a. & TESS & TESS \\
60283.9443 & 0.0011 & 9654 & -0.0002 & n.a. & TESS & TESS \\
60284.0132 & 0.0009 & 9655 & 0.0004 & n.a. & TESS & TESS \\
60284.0809 & 0.0010 & 9656 & -0.0004 & n.a. & TESS & TESS \\
60284.1499 & 0.0014 & 9657 & 0.0002 & n.a. & TESS & TESS \\
60284.2178 & 0.0007 & 9658 & -0.0003 & n.a. & TESS & TESS \\
60284.2865 & 0.0011 & 9659 & -0.0001 & n.a. & TESS & TESS \\
60284.3550 & 0.0014 & 9660 & 0.0000 & n.a. & TESS & TESS \\
60284.4240 & 0.0008 & 9661 & 0.0006 & n.a. & TESS & TESS \\
60284.4919 & 0.0007 & 9662 & 0.0001 & n.a. & TESS & TESS \\
60284.5603 & 0.0004 & 9663 & 0.0000 & n.a. & TESS & TESS \\
60284.6286 & 0.0006 & 9664 & -0.0001 & n.a. & TESS & TESS \\
60284.6968 & 0.0009 & 9665 & -0.0003 & n.a. & TESS & TESS \\
60284.7656 & 0.0015 & 9666 & 0.0002 & n.a. & TESS & TESS \\
60284.8341 & 0.0009 & 9667 & 0.0002 & n.a. & TESS & TESS \\
60284.9022 & 0.0010 & 9668 & -0.0001 & n.a. & TESS & TESS \\
60284.9710 & 0.0014 & 9669 & 0.0003 & n.a. & TESS & TESS \\
60285.0392 & 0.0011 & 9670 & 0.0000 & n.a. & TESS & TESS \\
60285.1076 & 0.0013 & 9671 & 0.0000 & n.a. & TESS & TESS \\
60285.1758 & 0.0008 & 9672 & -0.0002 & n.a. & TESS & TESS \\
60285.2446 & 0.0011 & 9673 & 0.0001 & n.a. & TESS & TESS \\
60285.3129 & 0.0011 & 9674 & 0.0001 & n.a. & TESS & TESS \\
60285.3812 & 0.0006 & 9675 & -0.0001 & n.a. & TESS & TESS \\
60285.4496 & 0.0010 & 9676 & -0.0001 & n.a. & TESS & TESS \\
60285.5179 & 0.0012 & 9677 & -0.0002 & n.a. & TESS & TESS \\
\hline
\end{longtable}


\begin{thebibliography}{}
\bibliographystyle{plainnat}


\bibitem[Andronov(1994)]{andronov1994}Andronov I.~L. 1994, {\it Odessa Astronomical Publications}, {\bf 7}, 49, \OEJVbibcode{1994OAP.....7...49A}

\bibitem[Andronov(2020)]{andronov2020}Andronov I.~L. 2020, {\it Knowledge Discovery in Big Data from Astronomy and Earth Observation}, 1st Edition. Edited by P. {\v{S}}koda and F. Adam. ISBN: 978-0-128-19154-5. Elsevier, 2020, p.191, \OEJVbibcode{2020kdbd.book..191A}

\bibitem[Andronov \& Baklanov(2004)]{andronov2004}Andronov, I.~L., \& Baklanov, A.~V. 2004 {\it Astronomical School's Report}, {\bf 5}, 264 \OEJVbibcode{2004AstSR...5..264A}

\bibitem[Andrych, Andronov, Chinarova \& Marsakova(2015)]{andrych2015} Andrych, K.~D., Andronov, I.~L., Chinarova, L.~L., \& Marsakova, V.~I. 2015, {\it Odessa Astronomical Publications}, {\bf 28}, 158, \OEJVbibcode{2015OAP....28..158A}

\bibitem[Andrych, Andronov \& Chinarova(2020)]{andrych2020}Andrych, K.~D., Andronov, I.~L., \& Chinarova, L.~L. 2020, {\it Journal of Physical Studies}, {\bf 24}, 1902, \OEJVbibcode{2020JPhSt..24.1902A}

\bibitem[Axelsen \& Napier-Munn(2020)]{axelsen2020}Axelsen, R.~A., \& Napier-Munn, T. 2020, {\it JAAVSO}, {\bf 48}, 241, \OEJVbibcode{2020JAVSO..48..241A}

\bibitem[Baglin et al.(1973)]{baglin1973}Baglin, A., Breger, M., Chevalier, C. et al. 1973, {\it A\&A}, {\bf 23}, 221, \OEJVbibcode{1973A\&A....23..221B}

\bibitem[Benn(2012)]{benn2012}Benn D., 2012, {\it JAAVSO}, {\bf 40}, 852, \OEJVbibcode{2012JAVSO..40..852B}

\bibitem[Benn et al.(2024)]{benn2024}Benn, D., Beck, S., Pyatnytskyy, M. et al. 2024, \OEJVlink{https://github.com/AAVSO/VStar}

\bibitem[Breger(2000a)]{breger2000a}Breger, M. 2000a, {\it Baltic Astronomy}, {\bf 9}, 149, \OEJVbibcode{2000BaltA...9..149B}

\bibitem[Breger(2000b)]{breger2000b}Breger, M. 2000b, {\it ASP Conference Series}, {\bf 210}, 3, \OEJVbibcode{2000ASPC..210....3B}

\bibitem[Butters et al.(2010)]{butters2010}Butters, O.~W., West, R.~G., Anderson, D.~R. et al. 2010, {\it A\&A}, {\bf 520}, L10, \OEJVbibcode{2010A\&A...520L..10B}

\bibitem[Collins et al.(2017)]{collins2017}Collins, K.~A., Kielkopf, J.~F., Stassun, K.~G., and Hessman, F.~V. 2017, {\it AJ}, {\bf 153}, 77, \OEJVbibcode{2017AJ....153...77C}

\bibitem[Drake et al.(2009)]{drake2009}Drake, A.~J., Djorgovski, S.~G., Mahabal, A. et al. 2009, {\it ApJ}, {\bf 696}, 870,  \OEJVbibcode{2009ApJ...696..870D}

\bibitem[Gaia Collaboration(2022)]{gaia2022}Gaia Collaboration 2022, {\it VizieR Online Data Catalog: Gaia DR3 Part 1. Main source}, \OEJVbibcode{2022yCat.1355....0G}

\bibitem[Gaia Collaboration(2023)]{gaia2023}Gaia Collaboration 2023, {\it A\&A}, {\bf 674}, id.A1, \OEJVbibcode{2023A\&A...674A...1G}

\bibitem[Gajdo{\v{s}} \& Parimucha(2019)]{gaidos2019}Gajdo{\v{s}}, P., \& Parimucha, {\v{S}}. 2019, {\it Astrophysics Source Code Library}, record ascl:1901.002, \OEJVbibcode{2019ascl.soft01002G}

\bibitem[Gajdo{\v{s}}(2023)]{gaidos2023}Gajdo{\v{s}}, P. 2023, {\it OEJV}, {\bf 241}, 1, \OEJVbibcode{2023OEJV..241....1G}

\bibitem[Handler(2009)]{handler2009}Handler, G. 2009, {\it AIP Conference Proceedings}, {\bf 1170}, 403, \OEJVbibcode{2009AIPC.1170..403H}

\bibitem[Henden et al.(2018)]{henden2018}Henden, A.~A., Levine, S., Terrell, D., Welch, D.~L., Munari, U., Kloppenborg, B.~K. 2018, {\it American Astronomical Society}, AAS Meeting \#232, id. 223.06  \OEJVbibcode{2018AAS...23222306H}

\bibitem[Kochanek et al.(2017)]{kochanek2017}Kochanek, C.~S., Shappee, B.~J., Stanek, K.~Z. et al. 2017, {\it PASP}, {\bf 129}, 104502, \OEJVbibcode{2017PASP..129j4502K}

\bibitem[Laney, Joner \& Rodriguez(2003)]{laney2003}Laney, C.~D., Joner, M., \&  Rodriguez, E. 2003, {\it ASP Conference Series}, {\bf 292}, 203, \OEJVbibcode{2003ASPC..292..203L}

\bibitem[Lightkurve Collaboration(2018)]{lightkurve2018}Lightkurve Collaboration, Cardoso, J.~V.~d.~M., Hedges, C. et al. 2018, {\it Astrophysics Source Code Library}, record ascl:1812.013, \OEJVbibcode{2018ascl.soft12013L}

\bibitem[Masci et al.(2019)]{masci2019}  Masci, F.~J., Laher, R.~R., Rusholme, B. et al. 2019, {\it PASP}, {\bf 131},  018003, \OEJVbibcode{2019PASP..131a8003M}

\bibitem[McNamara(1997)]{mcnamara1997}McNamara, D. 1997, {\it PASP}, {\bf 109}, 1221, \OEJVbibcode{1997PASP..109.1221M}

\bibitem[McNamara et al.(2000)]{mcnamara2000}McNamara, D.~H., Madsen, J.~B., Barnes, J., Erickson, B.~F. 2000, {\it PASP}, {\bf 112}, 202, \OEJVbibcode{2000PASP..112..202M}

\bibitem[McNamara(2011)]{mcnamara2011}McNamara, D.~H. 2011, {\it AJ}, {\bf 142}, 110,\OEJVbibcode{2011AJ....142..110M}

\bibitem[Pyatnytskyy(2021)]{pyatnytskyy2021}Pyatnytskyy, M. 2021, {\it JAAVSO}, {\bf 49}, 58, \OEJVbibcode{2021JAVSO..49...58P}

\bibitem[Pyatnytskyy \& Andronov(2024)]{pyatnytskyy2024}Pyatnytskyy, M.~Yu., \& Andronov, I.~L. 2024, {\it RNAAS}, {\bf 8}, 159 \OEJVbibcode{2024RNAAS...8..159P}

\bibitem[Ricker et al.(2014)]{ricker2014}Ricker, G.~R., Winn, J.~N., Vanderspek, R. et al. 2014, {\it Proceedings of the SPIE}, {\bf 9143}, 914320, \OEJVbibcode{2014SPIE.9143E..20R}

\bibitem[Stassun et al.(2019)]{stassun2019}Stassun, K.~G., Oelkers, R.~J., Paegert, M. et al. 2019, {\it AJ}, {\bf 158}, 138, \OEJVbibcode{2019AJ....158..138S}

\bibitem[Sterken(2005)]{sterken2005}Sterken C., ed. 2005, {\it ASP Conference Series}, {\bf 335}, \OEJVbibcode{2005ASPC..335.....S}

\bibitem[Watson, Henden \& Price(2006)]{watson2006}Watson, C.~L., Henden, A.~A., Price, A. 2006, The Society for Astronomical Sciences 25th Annual Symposium on Telescope Science. Held May 23-25, 2006, at Big Bear, CA. Published by the Society for Astronomical Sciences, p. 47 \OEJVbibcode{2006SASS...25...47W}

\bibitem[Xue et al.(2023)]{Xue2023}Xue, W., Niu, J.-S., Xue, H.-F., Yin, S. 2023, {\it Research in Astronomy and Astrophysics}, {\bf 23}, id.075002, \OEJVbibcode{2023RAA....23g5002X}

\bibitem[Ziaali et al.(2019)]{ziaali2019}Ziaali, E., Bedding, T.~R., Murphy, S.~J. et al. 2019, {\it MNRAS}, {\bf 486}, 4348, \OEJVbibcode{2019MNRAS.486.4348Z}

\end{thebibliography}
\end{document}